\documentclass[final]{IEEEtran} 
\usepackage{amsmath}
\usepackage{graphicx}
\usepackage[noadjust]{cite}

\begin{document}
\setlength{\parskip}{0pt}
\setlength{\dblfloatsep}{0pt}
\setlength{\floatsep}{0pt}
\setlength{\dbltextfloatsep}{0pt}
\setlength{\textfloatsep}{0pt}

\title{Non-volatile multilevel resistive switching memory cell: A transition metal oxide-based circuit} 

\author{P. Stoliar, P. Levy, M. J. S\'{a}nchez, A. G. Leyva, C. A. Albornoz, F.
Gomez-Marlasca, A. Zanini, C. Toro Salazar, N. Ghenzi, and M. J. Rozenberg%
\thanks{This work was partially supported by CONICET, ANPCyT, UNSAM, DuPont
and Fundaci\'{o}n Balseiro. M. J. S., M. J. R. and P. L. are members of
CONICET.

P. Stoliar is with the ECyT, Universidad Nacional de San Mart\'{i}n  
, Argentina, also with the GIA,
CAC-CNEA, Argentina, and also
with the Laboratoire de Physique des Solides, UMR8502 Universit\'{e} Paris-Sud
, France.

P. Levy, C. A. Albornoz, F. Gomez-Marlasca and N. Ghenzi
are with the GIA, CAC-CNEA,
Argentina.

M. J. S\'{a}nchez is with the Centro At\'{o}mico Bariloche and Instituto Balseiro,
CNEA, Argentina.

A. G. Leyva is with the GIA, CAC-CNEA (1650) San Mart\'{i}n,
Argentina, also with the ECyT, Universidad Nacional de San Mart\'{i}n, Argentina.

A. Zanini is with the Departamento de Ingenier\'{i}a Qu\'{i}mica, FIUBA, Universidad
de Buenos Aires, Argentina.

C. Toro Salazar is with the Departamento de micro y nano tecnolog\'{i}a,
CAC-CNEA, Argentina.

M. J. Rozenberg is with the Departamento de F\'{i}sica Juan Jos\'{e} Giambiagi,
FCEN, Universidad de Buenos Aires, Argentina, and also with the Laboratoire de Physique
des Solides, UMR8502 Universit\'{e} Paris-Sud, France.%

Copyright (c) 2013 IEEE. Personal use of this material is permitted.
However, permission to use this material for any other purposes must be
obtained from the IEEE by sending an email to pubs-permissions@ieee.org
}}

\maketitle

\begin{abstract}
We study the resistive switching (RS) mechanism as way to obtain multi-level
memory cell (MLC) devices. In a MLC more than one bit of information
can be stored in each cell. Here we identify one of the main conceptual
difficulties that prevented the implementation of RS-based MLCs. We
present a method to overcome these difficulties and to implement a
6-bit MLC device with a manganite-based RS device. This is done by
precisely setting the remnant resistance of the RS-device to an arbitrary
value. Our MLC system demonstrates that transition metal oxide non-volatile
memories may compete with the currently available MLCs. 
\end{abstract}


\section{Introduction}
During the last decade, the development of non-volatile electronic
memories based on the resistive switching (RS) effect in transition
metal oxides made a huge progress, becoming one of the promising
 candidates
to substitute the standard technologies in a near future. RS refers
to the reversible change of the resistance of a nanometer-sized media
by the application of electrical pulses
\cite{Sawa2008,Rozenberg2010,ADMA:ADMA200900375,Strukov2008,itrs2011,lee2008low}.

Though immediately after the discovery of the reversible RS effect
in transition metal oxides its application for multi-level cell memories
(MLC) was envisioned \cite{Beck2000,Alibart2012,Kim2010,He2012,schindler2007bipolar},
reports so far chiefly focused on single-level cell (SLC) devices
\cite{Sawa2008,ADMA:ADMA200900375}. Unlike SLC, which can only store
one-bit per cell, MLC memories may store multiple bits in a single
cell \cite{Ricco1998}.
MLCs of up to 4 bits (16 levels) are currently available based in both, 
standard Flash \cite{Tanakamaru2012} and phase-change \cite{Papandreou2011} technologies.”

\begin{figure}
\begin{center}
\includegraphics[width=6.0cm]{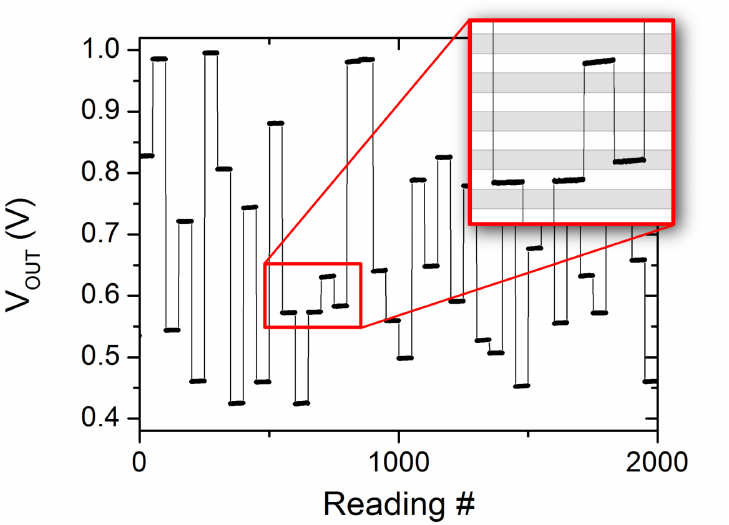}
\caption{Experimental Results. Random memory
level sequence stored in the 6-bit MLC. The plot presents the memory
states during 50 readings (at 10 Hz) for each stored value. For a
6-bit MLC, the working memory range is divided into 64 levels (as
indicated in the inset).}
\label{fig:zoom}
\end{center}
\end{figure}

In an $n$-bit RS-based MLC memory, one has to encode the $2^{n}$
memory levels as distinct resistance states. 
Thus, the $2^{n}$ memory states can be identified with each one of
the consecutive and adjacent resistance \lq\lq{}bins\rq\rq{}
of width $\triangle\mathcal{R}=\mathcal{R}_{i+1}-\mathcal{R}_{i}$, with $i=1,2,..2^{n}$. To store
the $i^{th}$ memory state 
the device has to be \lq\lq{}SET\rq\rq{} to
the corresponding bin, ie $\mathcal{R}_{i}<\mathcal{R}<\mathcal{R}_{i+1}$.
The requirement of non-volatility implies that the 
value should remain within that bin, even after the input is disconnected
or the memory state is read out. 
In Fig. \ref{fig:zoom} we plot data
of a random sequence of stored memory states in our implementation
of a 64-level (6-bit) MLC.
For practical convenience
resistance levels are translated into voltage levels by a fixed bias current.

The central component of the MLC is a RS device of a resistance $\mathcal{R}$,
whose magnitude can be changed by application of a current pulse $I_{pulse}$.
The resistance change depends on $I_{pulse}$ through a highly non-trivial
function $f$; $\triangle\mathcal{R}=f\left(I_{pulse}\right).$ The
function $f$ is usually unknown, however, an important general requirement
for non-volatile memory applications using RS is that $f=0$ for currents
below a given threshold. This allows to sense (ie, read-out) a stored
memory state, ie the so called remnant resistance $\mathcal{R}_{rem}$,
injecting a bias current $\left|I_{0}\right|<\left|I_{th}\right|$
without modifying the stored information. Thus, $\mathcal{R}_{rem}\equiv\mathcal{R}\left(I_{0}\right)$,
$\left|I_{0}\right|<\left|I_{th}\right|$. A two level memory can
be simply implemented by pulsing a RS-device strongly with opposite
polarities and sensing the corresponding high and low $\mathcal{R}$
states with a weak bias current. However, the previous observation
of a current threshold has an important consequence for the implementation
of MLC memories. In fact, in order to implement an MLC one should
be able to tune the value of $\mathcal{R}_{rem}$ to any desired value.
The better the control on this tuning, the better the ability to define
the \lq\lq{}bins\rq\rq{} and a larger number of
bits could be coded in the MLC. In principle, a perfect knowledge
of the function $f$ should allow for this, but in practice the function
is not known. Yet, any reasonable form of $f$ would allow to tune
any arbitrary memory state by applying a sequence of pulses of decreasing
intensity, following a simple \lq\lq{}zero-finding\rq\rq{} 
algorithm as in standard control circuit. Nevertheless, the requirement
of a threshold current difficult the fine tuning, as small corrections
beneath $|I_{th}|$ have no effect. 
In practice, the dead-zone introduced by relatively large values of  $|I_{th}|$ prevents 
a straightforward implementation of RS-device as MLC memories with a large
number of levels. Below we shall demonstrate how this conceptual problem
can be overcome, and exhibit the implementation of a 64-level MLC.

\section{Implementation}

\begin{figure}
\begin{center}
\includegraphics{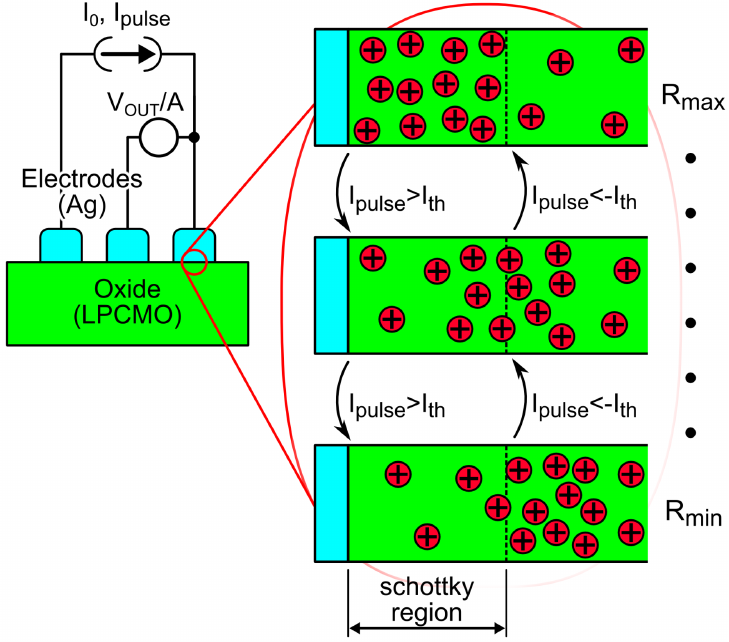}
\caption{Left: Three-contact setup of
the RS-device. Right: Schematic representation of the physical mechanism
of resistive switching in Ag/LPCMO. The current pulses progressively
change the profile of the oxygen vacancies within a nanometer-sized
region in proximity to the Ag contacts. We define positive pulses
those flowing from the electrode towards the LPCMO.}
\label{phys}
\end{center}
\end{figure}

\begin{figure}
\begin{center}
\includegraphics{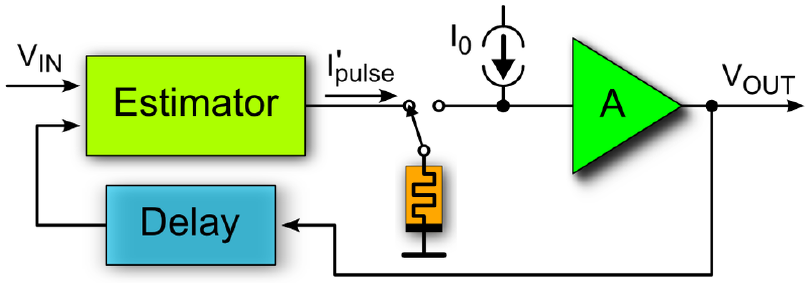}
\caption{Block diagram of the implementation.
The switch commutes the system between a read state, where the resistance
of the RS-device is sense with $I_{0}$, and a write state where current
pulses are applied in order to cancel the difference between $V_{IN}$
and $V_{OUT}$. When WR is deactivated, $I_{pulse}$ is force to zero.}
\label{F4}
\end{center}
\end{figure}

We adopt a manganite-based RS-device, made by depositing silver
contacts on a sintered pellet of 
La$_{0.325}$Pr$_{0.3}$Ca$_{0.375}$MnO$_{3}$
(LPCMO) \cite{Ghenzi2010}. 
A RS-device is defined between the
pulsed Ag/LPCMO and a second non-pulsed contact. A third electrode
(earth) is  required in a minimal 3-contact configuration setup.

A requirement for our MLC is that the RS-device operates in bipolar
RS mode \cite{Sawa2008,Rozenberg2010,ADMA:ADMA200900375}, ie, depending
on the pulse polarity the remnant resistance either increases or decreases.

It is now well established that the mechanism behind
 bipolar  RS is the redistribution of oxygen vacancies, within the nanometer-scale
region of the sample in contact with the electrodes \cite{Rozenberg2010}.
 In the case of LPCMO, the  oxygen vacancies  significantly
increase its resistivity because the electrical transport relies
on the double-exchange mechanism mediated by the oxygen atoms \cite{Dagotto2003}.

 Pulsing an electrical current trough the contact
will produce a large electric field at the interface due to the high
resistivity of the Ag/LPCMO Schottky barrier. If the pulse is strong
enough, it will enable the migration oxygen ions across the barrier, modifying the concentration of vacancies and, hence,
changing the interface resistance. 
 The ionic migration remains always near the interface and does not penetrate
deep into the bulk, since the much larger conductivity there prevents
the development of high electric fields. Thus, the RS effect remains
confined to a nanometer-sized region near the interface, as schematically
depicted in Fig. \ref{phys}. 

We now introduce a practical implementation of an RS-based MLC that
produced the results shown in Fig. \ref{fig:zoom}. We used an Ag/LPCMO
interface and off-the-shelf electronic components. The block diagram
of the concept is presented in Fig. \ref{F4} and the schematics and
technical details are included in the Appendix. We envisage an implementation
of an RS-based MLC memory chip where the storage core is a set of
many RS-units with a single common control circuit. The common control
would set each of the individual RS-units, one at a time, thus resembling
the concept of the refresh logic circuit in the DRAMs. 
In this work we
demonstrate the implementation of the control circuit with a single
RS memory unit.

Key to our MLC implementation is to adopt a discrete-time 
algorithm that overcomes the problems discussed above \cite{Ogata1995},
and which we describe next. The required memory state $\mathcal{R}_{i}$
is coded in terms of a $V_{IN}$, while $V_{OUT}$ indicates the actual
stored value (Fig. \ref{F4}). The system iteratively
applies pulses $I_{pulse}$ of a strength that is an estimate of the
required value to set the target state $V_{IN}$, eventually converging
to it. This discrete-time feedback loop continuously cycles between
2 stages; \lq\lq{}probe\rq\rq{} and \lq\lq{}correct\rq\rq{}.
In probe stage the switch connects the RS-device to the current source
$I_{0}$ in order to sense the remnant resistance. In the correct
stage, the switch connects the RS-device to a pulse generator that
applies a corrective pulse $I_{pulse}$, of a strength that is obtained
from the difference between the delayed $V_{OUT}$ and the target
value $V_{IN}$ as follows,
\begin{equation}
I_{pulse}\left[k\right]=K_{P}\,e\left[k\right]+K_{I}\,\sum_{i=0}^{k}e\left[i\right],\label{eq:PI}
\end{equation}
 where the error signal $e\left[k\right]=V_{OUT}\left[k-1\right]-V_{IN}\left[k\right]$ 
is the change required in the output
voltage and $I_{pulse}\left[k\right]:=I_{pulse}\left(k\, f_{{\rm CLK}}^{-1}\right)$, 
being $f_{{\rm CLK}}$ the frequency of the system clock and $k$ an integer.
 $K_{P}$ and $K_{I}$ are generic proportional and integral constants respectively with $\frac AV$ unit, being $A$ and $V$ the electric current and voltage units.
The first term of the equation represents a proportional
estimator. The second term 
prevents the system to get stuck in a condition where $\left|I_{pulse}\right|<\left|I_{th}\right|$.
In fact, for low $e\left[k\right]$ a pulse of strength $K_{P}\,e\left[k-1\right]$
would lay below the threshold, thus not producing any further change
in the state of the system. The magnitude of the second term linearly
increases in time, thus making $I_{pulse}$ to eventually overcome
the threshold and correct the output voltage in the desired direction.

Notice that even if this approach resembles a standard proportional-integral  (PI)
control loop with dead-zone, there are substantial differences. First,
the remnant resistance reading and the correcting pulse application
occur at different times. This requires the addition of the continuously
commuting switch. Second, we also needed the introduction of a delay,
implemented as a sample and hold circuit, as is required for the feedback
path. These differences and the strong non-linearity in $f$ makes
the stability analysis of this approach (which also depends on specific
values of $K_{P}$ and $K_{I}$) a significant issue that we describe below.

\section{Stability analysis}
\subsection{Discrete-Time Model}

\begin{figure}
\begin{center}
\includegraphics{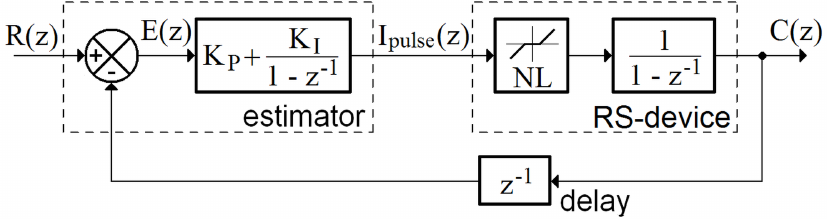} 
\caption{Discrete-time model. In each cycle the input signal
$R\left(z\right)$ is compared with the output signal $C\left(z\right)$
after being delayed, generating the error signal $E\left(z\right)$.
Based on this signal, the estimator generates the corrective pulses
$I_{pulse}\left(z\right)$.}
\label{fig:concept}
\end{center}
\end{figure}

The study of the system stability is based on the discrete-time model
presented in Fig. \ref{fig:concept}.
In the z-domain, Eq. \ref{eq:PI} becomes
\begin{equation}
I_{pulse}\left(z\right)=E\left(z\right)\,\left(K_{P}+\frac{K_{I}}{1-z^{-1}}\right).\label{eq:PI k-domain}
\end{equation}

The resulting $\mathcal{R}_{rem}$ after these corrective pulses is probed
by connecting the RS-device to the bias current source ($I_{0}$),
obtaining the output signal $c\left[k\right]$($C\left(z\right)$ in z-domain). In the next cycle,
it will be compared to the reference input $r\left[k+1\right]$ generating
the error signal $e\left[k+1\right]$. This fact implies
the 1-cycle delay $z^{-1}$.

Central to the present proposal is the following equation, a discrete-time
model for the RS-device: 
\begin{equation}
\mathcal{R}_{rem}\left[k\right]=R_{0}+R_{1}\,\sum_{j=-\infty}^{k}\mathrm{NL}\left(I_{pulse}\left[j\right]\right),\label{eq:rrem}
\end{equation}
 where $R_{0}$ and $R_{1}$, having resistance units, are an offset
and a proportionality factor respectively. $\mathrm{NL}$ is a non-linear
function that has to be defined on the basis of the general behavior
of a RS-device operating in bipolar mode.
In this model the $\mathcal{R}_{rem}$ is calculated by integrating
the writing pulses after been weighted by the $\mathrm{NL}$ function.
In this way, we model the possible change of the device resistance after the $k$-th  pulse as $\triangle\mathcal{R}_{rem}\left[k\right]=\mathcal{R}_{rem}\left[k\right]-\mathcal{R}_{rem}\left[k-1\right]=
R_{1}\mathrm{NL}\left(I_{pulse}\left[k\right]\right)$.

A concrete implementation of NL is presented below 
\begin{equation}
\mathrm{NL}\left[k\right]=\begin{cases}
\left(I_{pulse}\left[k\right]+I_{th}\right)\frac1A & I_{pulse}\left[k\right]<-I_{th}\\
0 & -I_{th}<I_{pulse}\left[k\right]<I_{th}\\
\left(I_{pulse}\left[k\right]-I_{th}\right)\,u_{1} &  I_{pulse}\left[k\right]>I_{th}\,.
\end{cases}\label{eq:nl}
\end{equation}
The unit of $u_{1}$ is $\frac1A$.When a negative
signal exceeds the threshold $I_{pulse}\left[k\right]<-I_{th}$, $\mathcal{R}_{rem}$
decreases as a linear function of $I_{pulse}\left[k\right]$ with slope 1$\frac \Omega A$.
For a positive signal greater than $I_{th}$, $\mathcal{R}_{rem}$
increases with slope $u_{1}\Omega$. For the sake of stability analysis
Eq. \ref{eq:rrem} is simplified as:
\begin{equation}
c\left[k\right]=V\sum_{j=-\infty}^{k}\mathrm{NL}\left(I_{pulse}\left[j\right]\right).\label{eq:c k-domain}
\end{equation}
In this way, the analysis is not considering any instability when sensing $\mathcal{R}_{rem}$
with $I_0$. Indeed, our actual implementation did not present any critical issue at this level 
(see the Appendix).
From now on, unit-step sequence is considered as input signal \cite{Chen1993}.
The steady-state error for the system with $K_{I}=0$ (just proportional
control) and $u_{1}=1A^{-1}$ is 
$e_{s}=\lim_{k\rightarrow\infty}e\left[k\right]=\frac{I_{th}}{K_{P}}$,
then the steady-state response is $c_{s}=1-\frac{I_{th}}{K_{P}}$.
In fact, Fig. \ref{fig:PI}(a) shows that the system converges to the required
set-point only for $I_{th}=0$. The system arrives to this condition
because when $\left|e\left[k\right]\right|\leq\frac{I_{th}}{K_{P}}$
the excitation of the RS-device is $\left|I\left[k\right]\right|\leq I_{th}$,
i.e. lower than the minimum voltage required to produce a change in
$R_{rem}$. A proportional control that enters into this condition,
remains there indefinitely. 
The integral term in Eq. \ref{eq:PI k-domain}, that turns the system
into a PI control, avoids the problem; when continuously
integrated, $e\left[k\right]$ makes $I_{pulse}\left[k\right]$ to eventually
overcome the threshold $\left|I_{th}\right|$. See Fig. \ref{fig:PI}(b) and (c).

\subsection{Analysis without $\mathrm{NL}$ }%%%%%%%%%%%%%%%%%%%%
We begin by considering the simplified situation
where  the  stability is analyzed by removing the nonlinearities introduced by
$\mathrm{NL}\left[k\right]$ ($u_{1}=1A^{-1}$ and $I_{th}=0$). The
transfer function of the system is 
$\frac{C\left(z\right)}{R\left(z\right)}=\frac{K_{P}+K_{I}-K_{P}\,z^{-1}}{1+\left(K_{P}+K_{I}-2\right)\,z^{-1}+\left(1-K_{P}\right)\,z^{-2}}$.
For $K_{I}=0.25\frac AV$ (typical value) the system is critically damped when
$K_{P}=0.75\frac AV$, and remains stable for $K_{P}<1.875\frac AV$.
Increasing $K_{I}$ moves the location of the poles following $\left|z\right|=1$,
arriving to $z=-1$ when $K_{I}=4\frac AV$, where the system
becomes unstable for any $K_{P}$ (see Fig. 5 (m)).

\subsection{Analysis with $\mathrm{NL}$}%%%%%%%%%%%%%%%%%%%%
Although the introduction of nonlinearities in Eq. \ref{eq:nl} does
not allow to study the system analytically, as in the previous section,
its response might be numerically simulated.
In fact, simulations where performed in the $k$-space by solving Eq. 5, after substituting $I_{pulse}\left[k\right]$ (Eq. 1) and $\mathrm{NL}\left[k\right]$ (Eq. 4).
For the computation of $e\left[k\right]$, we assigned  $V_{OUT}\left[k\right]=c\left[k\right]$ and the unitary step function at the input (ie. $V_{IN}\left[k\ge0\right]=1V$).

\begin{figure*}
\begin{center}
\includegraphics{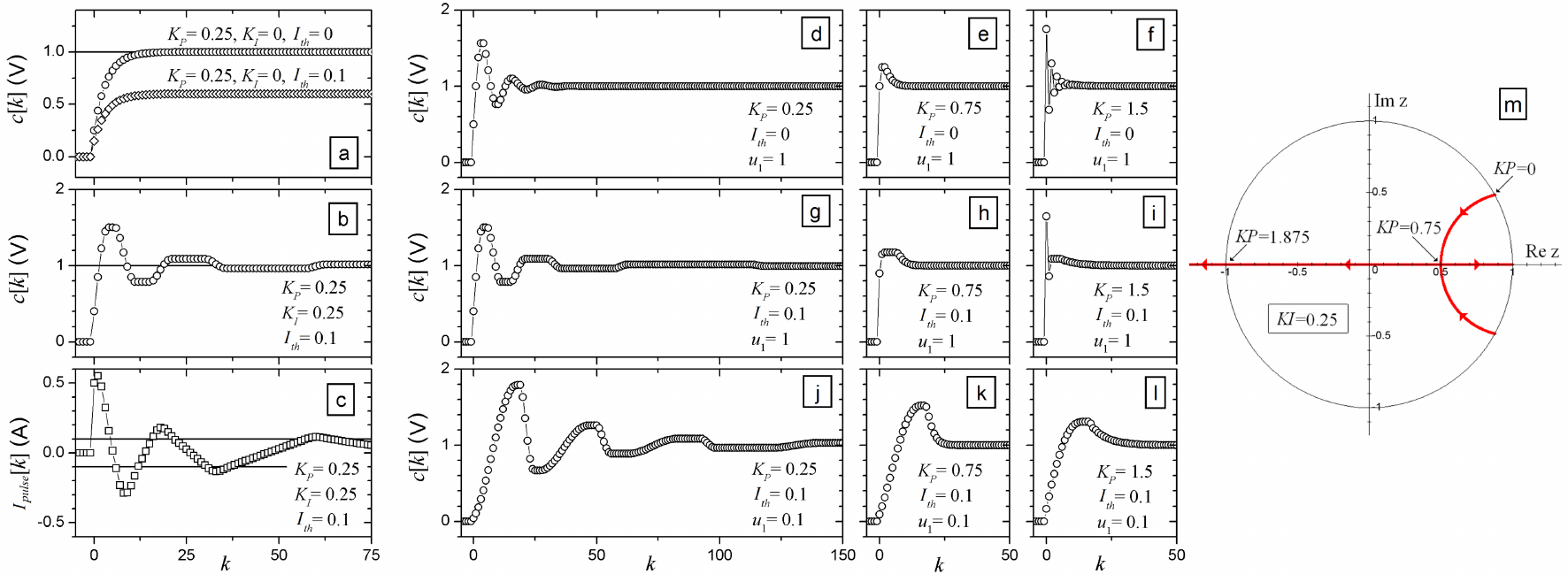} 
\caption{Simulated output $c\left[k\right]$ of the system upon
an unitary-step input for a proportional control (a) and for a proportional-integral
control loop (b). (c) is the excitation $I_{pulse}\left[k\right]$ applied
to the RS-device model corresponding to (b). $c\left[k\right]$ does
not change when $\left|I_{pulse}\left[k\right]\right|\leq0.1A$. $u_{1}=1A^{-1}$
in all these plots. 
Panels d-l present different conditions for non-zero $K_{P}$ (indicated in the panels), and $K_{I}=0.25$. 
(m) Root locus for the system without $\mathrm{NL}$.
$K_{I}$ and $K_{P}$ are indicated in $\frac AV$
units and $I_{th}$ in Amperes.}
\label{fig:PI}
\end{center}
\end{figure*}

Fig. \ref{fig:PI} (d-f) shows this simulated response
of the system for three sets $\left\{ K_{P},K_{I}\right\} $
corresponding to representative position of the poles 
of the equivalent system without nonlinearities.

Fig. \ref{fig:PI} (g-i) shows the effect of introducing the
threshold ($I_{th}\neq0$) for the same set $\left\{ K_{P},K_{I}\right\} $.
The system response is qualitatively the same with the addition of
periods where $c\left[k\right]$ is ``frozen'' because $\left|I_{pulse}\left[k\right]\right|\leq I_{th}$,
while the integral term is growing.
Fig. \ref{fig:PI} (j-l) shows the effect of further introducing
asymmetry into the nonlinear function $u_{1}\neq1A^{-1}$. The
effect that this nonlinearity introduces when $u_{1}<1A^{-1}$ is
equivalent to reducing the gain of the system for $I_{pulse}\left[k\right]>0$
(see Eq. \ref{eq:nl}). In fact, Fig. \ref{fig:PI}(j) evidences
a clear difference between the system speed when $c\left[k\right]$
is increasing as compared to the speed when $c\left[k\right]$ is
decreasing. 
The case for $u_{1}>1A^{-1}$ is equivalent to a case where 
$u_{1}\rightarrow u_{1}^{-1}$ and $K_{P}\rightarrow K_{P}\,u_{1}^{-1}$.

Simulations also show a change in the stability limit as reported
in the following table:

\vspace{2 mm}

%\begin{table}[!ht]
\renewcommand{\arraystretch}{1.3}
%\centering
%\begin{center}
{\centering
\begin{tabular}{ccc}
\hline 
$I_{th}\;(A)$  & $u_{1}\;(\frac1A)$  & stability limit ($K_{I}=0.25$)\tabularnewline
\hline 
0  & 1  & $K_{P}=1.875$\tabularnewline
0.1  & 1  & $K_{P}=1.969$\tabularnewline
0.1  & 0.1  & $K_{P}=11.1181$\tabularnewline
\hline 
\end{tabular}
%\end{center}
%\end{table}
%

}

\vspace{2 mm}

Summarizing, the simulations show that the system stability is not
compromised after the introduction of nonlinearities, even if in some
conditions, the system might require a considerably higher number
of cycles to stabilize.

\section{Results}

\begin{figure}
\begin{center}
\includegraphics{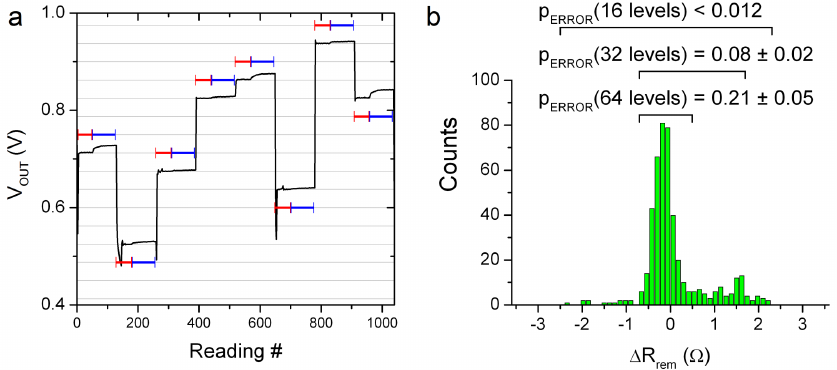}
\caption{Retentivity of the LPCMO-based
MLC. (a), $V_{OUT}$ continuously sampled at 10 reading per second.
During the readings indicated with the red bars the signal WR was
active; the output value accurately follows the sequence of input
values that is randomly chosen on the grid. During the readings indicated
with blue bars, no pulses were applied to the RS-device, resulting
in a drift. (b) Histogram of the last read value of the memory in
each cycle (ie, last value of blue sections). The horizontal bars
denote the $\Delta{R}$ intervals corresponding to a 16, 32 and 64
level devices. The location of the bars are chosen so to minimize
the error probability (ie, area of histogram outside the interval).}
\label{results}
\end{center}
\end{figure}

Two tests were done to evaluate the performance of the MLC memory.
Experimental results for the memory retention test are presented in
Fig. \ref{results}. We emphasize that the relatively slow operation
speed of our memory MLC would be dramatically improved upon device
miniaturization and integration. For simplicity, we sampled a random
subset of 16 out of 64 different voltages uniformly distributed along
the operative range (Fig. \ref{results}(a)). Each write and read cycle
had a duration of 13 seconds. In the first 5 s the WR signal was active
and a memory value was SET. After a memory value was SET, we observed
a small relaxation of $V_{OUT}$ with time constant of $\tau\sim$1.6
s. The value of $V_{OUT}$ remained essentially stable afterwards.
Thus, for the sake of performing a large number measurements, we only
monitored the memory retentivity during the 5 time-constants (ie,
8 seconds) that immediately followed the SET. Within that period,
the WR signal was inactive, and the memory state stored in $V_{OUT}$
was probed every 0.1 s (ie, the memory was read out 80 times).

In Fig. \ref{results}(b) we present the histogram of the distribution
of observed errors in the reading of a given stored memory value.
The observed values correspond to the remnant resistance of the memory
cell, after reading it 80 times at a 10 Hz rate, following the SET.
The histogram is constructed from a sequence of 460 memory state recordings.

The finite dispersion of the histogram is the main limitation for
implementing a high number of memory levels. To easily visualize the
retentivity performance of the memory for increasing number of levels,
we indicate with horizontal bars the resistance interval $\Delta R$
( =$\Delta V/I_{0}$) that corresponds to having a 16, 32 and 64 MLC
memory (ie, 4, 5 and 6 bits, respectively). The probability of error
in storing a memory state corresponds to the area of the histogram
that lays outside the resistance interval normalized to the total
area. With a confidence level of 95\%, the probability that the MLC
fails to retain a stored state is 0.21$\pm$0.05 for a 64-level system
(6 bit MLC). This error probability rapidly improves as the number
of levels is decreased, being $<$ 0.012 for a 16-level MLC. 

The previous data on the performance of the memory can also be expressed
in terms of the bit error rates (BER).
In our MLC, we find BER$\leq$0.006
for the 16-level system, BER$\leq$0.07 for the 32-level system and
BER$\leq$0.1 for the 64-level system. Lower error probabilities for
bits than for levels, reflects the fact that, typically, the error
corresponds to the memory level drifting to an immediate neighbor
level (that often shares a large number of bits).

\section{Conclusions}

In conclusion, we have introduced a  feedback algorithm
to precisely set the remnant resistance of a RS-device to an arbitrary
desired value within the device working range. This overcomes the
conceptual problem of fine tuning a resistance value in non-volatile
RS-devices for MLC applications, due the presence of a threshold current
behavior. The feedback configuration is intrinsically time-discrete,
since it is based on read / write sequences. The overhead in the writing
time introduced in this feedback system may be a limitation for its
utilization as primary memory in a computer system, but it can easily
compete with actual speed of the current mass storage devices (flash
memories). The applicability of the concept to implement an n-bit
MLC memory was successfully demonstrated for $n$ = 4, 5 and 6, with
an LPCMO-based circuit; which clearly illustrates the critical trade-off
between BER, number of memory levels and (power and area) overhead
of the control circuitry.

\appendix[Schematics and technical details of the circuit]%%%%%%%%%%%%%%%%%%%%

\begin{figure}
\begin{center}
\includegraphics{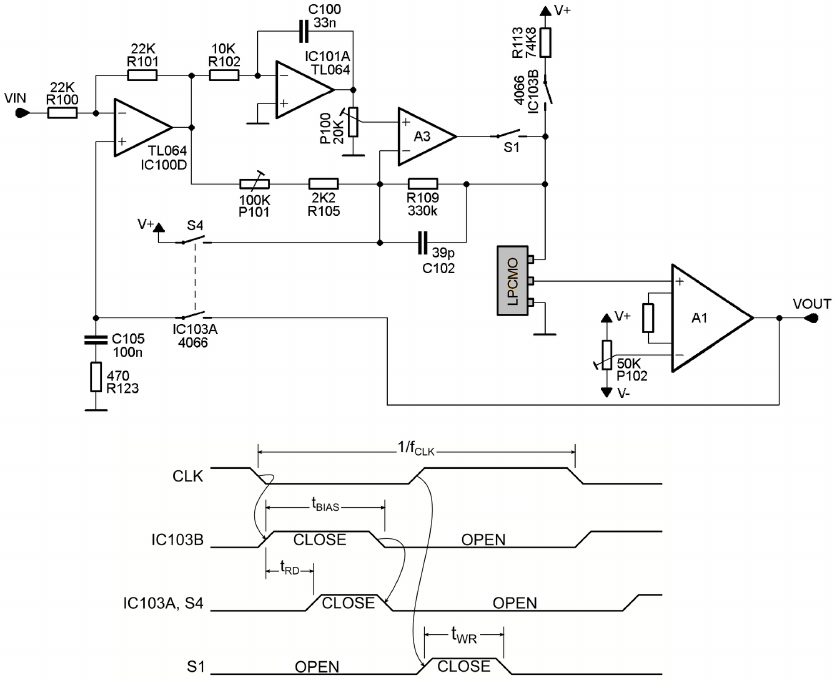} 
\caption{Simplified circuit of
the proposed implementation with its corresponding timing diagram.
In this proof-of-concept implementation $f_{{\rm CLK}}=$33Hz,
$t_{{\rm BIAS}}=1.7$ms, $t_{{\rm RD}}=0.25$ms and $t_{{\rm WR}}=120\mu$s.}
\label{fig:Simplified-circuit}
\end{center}
\end{figure}

Fig. \ref{fig:Simplified-circuit} shows a simplified 
circuit of our implementation.
The \lq\lq{}estimator\rq\rq{} circuit (see Fig. \ref{F4} and Fig. \ref{fig:concept}) 
is implemented by IC100D, IC101A and the high-current buffer A3 (and associated components).
It compares the output of the \lq\lq{}sample and hold\rq\rq{}  (S\&H) 
against the input signal divided by 2, and then computes the PI function. 
$K_{P}$ and $K_{I}$ are set by means of P101 and P100 respectively.

During the \lq\lq{}correct\rq\rq{} state, 
corrective 
pulses are applied to the
RS-devices (LPCMO), by closing S1 during
  $t_{{\rm WR}}=120\mu$s. C102 and
R109 introduce a time constant of $\sim13\mu$s, reducing the rise
time of the pulses in order to avoid overshoots. 
In the \lq\lq{}probe\rq\rq{} state, IC103B closes first and 
 $t_{{\rm RD}}=0.25$ms after, IC103A. Also, S4
clamps the inverting input of A3 to $\sim$V+. 
The timing of the circuit is generated by a timing circuit based on the master clock CLK  (see Fig. 7).
A current 
$I_{0}\approx\frac{V+}{R113\parallel R109}\approx120\mu A$
flows trough the LPCMO (V+ = 7.5V, V- = -7.5V). The voltage drop
at the switching interface of the RS-device (ranging 20-50 mV) is amplified by the instrumentation
amplifier A1 (gain=21), eventually setting the voltage at the output of the
S\&H. 
Both interfaces of the RS-device behave complementary 
(ie. when one interface reduces its resistance, the other increases \cite{Rozenberg2010}), 
then the total drop across the device is $\sim$70 mV 
and quite insensitive to the state, and therefore, $I_0$.

%\bibliographystyle{ieeetr}
%\bibliography{IEEE2}

\end{document}